\documentclass[]{spie}  
\usepackage[]{graphicx}
\usepackage{bm}
\usepackage{amsmath,amssymb,epsfig} 
\usepackage{wrapfig}

\title{A Search for New Physics with the BEACON Mission} 

\author{Slava G. Turyshev\supit{a}, Benjamin Lane\supit{b}, Michael Shao\supit{a}, and Andr\'e Girerd\supit{a}
\skiplinehalf
\supit{a}Jet Propulsion Laboratory, California Institute of Technology,\\
4800 Oak Grove Drive, Pasadena, CA 91109-0899, USA; \\
\supit{b}The Charles Stark Draper Laboratory, Inc. \\
555 Technology Square
Cambridge, MA 02139-3563, USA
}

\authorinfo{Further author information: turyshev@jpl.nasa.gov}

\begin{document} 
  \maketitle 

\begin{abstract}
 The primary objective of the Beyond
  Einstein Advanced Coherent Optical Network (BEACON) mission is a search for
  new physics beyond general relativity by measuring the curvature of   
  relativistic space-time around Earth. This curvature is characterized by 
  the Eddington parameter $\gamma$ -- the most
  fundamental relativistic gravity parameter and a direct measure
  for the presence of new physical interactions.  
 BEACON will
  achieve an accuracy of $1\times10^{-9}$ in measuring the
  parameter $\gamma$, thereby going a factor of 30,000 beyond the present
  best result involving the Cassini spacecraft.
  Secondary mission objectives include: (i) a direct measurement of
  the ``frame-dragging'' and geodetic precessions in the Earth's
  rotational gravitomagnetic field, to 0.05\% and 0.03\% accuracy
  correspondingly, (ii) first measurement of gravity's non-linear
  effects on light and corresponding 2nd order spatial metric's effects to
  0.01\% accuracy.    
BEACON will lead to robust advances in tests of fundamental physics -- this mission could discover a violation or extension of general relativity and/or reveal the presence of an additional long range interaction in physics. BEACON will provide crucial information to separate modern scalar-tensor theories of gravity from general relativity, probe possible ways for gravity quantization, and test modern theories of cosmological evolution.

\end{abstract}



\section{INTRODUCTION}
\label{sec:intro}  

After almost ninety years since general relativity was born, Einstein's general theory of relativity has survived every test \cite{Turyshev:2008jd,Will-lrr-2006-3}. Such longevity, of course, does not mean that this theory is absolutely correct, but it serves to motivate more accurate tests to determine the level of accuracy at which it is violated. General  relativity began its empirical success in 1915 by explaining the anomalous perihelion precession of Mercury's orbit, using no adjustable theoretical parameters.  Shortly thereafter, Eddington's 1919 observations of star lines-of-sight during a solar eclipse confirmed the doubling of the deflection angles predicted by the theory, as compared to Newtonian-like and Equivalence Principle arguments.  This test made  general relativity an instant success.\cite{Turyshev:2008jd} 

From these beginnings, the general theory of relativity has been verified at ever higher accuracy. Microwave ranging to the Viking landers on Mars yielded an accuracy of $\sim$0.2\% in the tests of general relativity \cite{Shapiro-etal:1976,Reasenberg-etal:1979}.  Astrometric observations of quasars on the solar background performed with Very-Long Baseline Interferometry (VLBI) improved the accuracy of tests of gravity to $\sim$0.045\% \cite{Shapiro_SS_etal:2004}. Lunar laser ranging (LLR) verified general relativity to $\sim$0.011\% via precision measurements of the lunar orbit \cite{Williams-etal:2004}.  These efforts culminated in 2003 with a solar conjunction experiment involving the Cassini spacecraft that improved the accuracy of the tests to $\sim$0.0023\% \cite{Bertotti-etal:2003}.

However, there are important reasons to question the validity of Einstein's theory of gravity.  Despite the beauty and simplicity of general relativity, our present understanding of the fundamental laws of physics has several shortcomings. The continued inability to merge gravity with quantum mechanics \cite{Turyshev-etal:2007}, and recent cosmological observations that lead to the unexpected discovery of the accelerated expansion of the universe (i.e., ``dark energy'') indicate that the pure tensor gravity field of general relativity needs modification.  It is now believed that new physics is needed to resolve these issues \cite{Turyshev-etal:2007}. Given the immense challenge posed by the unexpected discovery of the accelerated expansion of the universe, it is important to explore every option to explain and probe the underlying physics. Theoretical models of the kinds of new physics that can solve the problems above typically involve new interactions, some of which could manifest themselves as violations of the equivalence principle, variation of fundamental constants, modification of the inverse square law of gravity at various distances, Lorenz-symmetry breaking, large-scale gravitational phenomena, and corrections to the curvature of space-time around massive bodies (see disucssion in Ref.~\citenum{Turyshev-etal:2007,Turyshev:2008jd}).    

The Eddington parameter $\gamma$, whose value in general relativity is unity, is perhaps the most fundamental PPN parameter \cite{Will-lrr-2006-3}, in that $\frac{1}{2}(\gamma-1)$ is a measure, for example, of the fractional strength of the scalar gravity interaction in scalar-tensor theories of gravity \cite{Turyshev:2008jd}.  Currently, the most precise value for this parameter, $\gamma -1 = (2.1\pm2.3)\times10^{-5}$, was obtained using radio-metric tracking data received from the Cassini spacecraft \cite{Bertotti-etal:2003} during a solar conjunction experiment. This accuracy approaches the region where multiple tensor-scalar gravity models, consistent with the recent cosmological observations, predict a lower bound for the present value of this parameter at the level of $\gamma-1 \sim 10^{-6}-10^{-7}$ (see discussion in Ref.~\citenum{Turyshev-etal:2007,Turyshev:2008jd} and references therein). Therefore, improving the measurement of this parameter would provide crucial information to separate modern scalar-tensor theories of gravity from general relativity, probe possible ways for gravity quantization, and test modern theories of cosmological evolution.

The Beyond Einstein Advanced Coherent Optical Network (BEACON) concept is an astrophysics experiment designed to test the metric nature of gravitation -- a fundamental postulate of Einstein's general theory of relativity.  BEACON's primary mission objective is a precise measurement of the curvature of space-time characterized by the Eddington parameterized post-Newtonian parameter $\gamma$, which is regarded as a direct measure for the presence of a new physical interaction.  BEACON will achieve an accuracy of a part in 10$^9$ in measuring the parameter $\gamma$, i.e. a factor of 30,000 beyond the present best result of $\gamma-1=(2.1\pm2.3)\times10^{-5}$ provided by microwave tracking of the Cassini spacecraft on its approach to Saturn.\cite{Bertotti-etal:2003} The proposed BEACON development will rely on 20 years of experience in designing and building laser metrology systems for astrophysics.  The effort will also benefit from access to SIM technology and the gravitational research community.  

\begin{wrapfigure}{R}{0.420\textwidth}
  \vspace{-25pt}
  \begin{center}
    \includegraphics[width=0.420\textwidth]{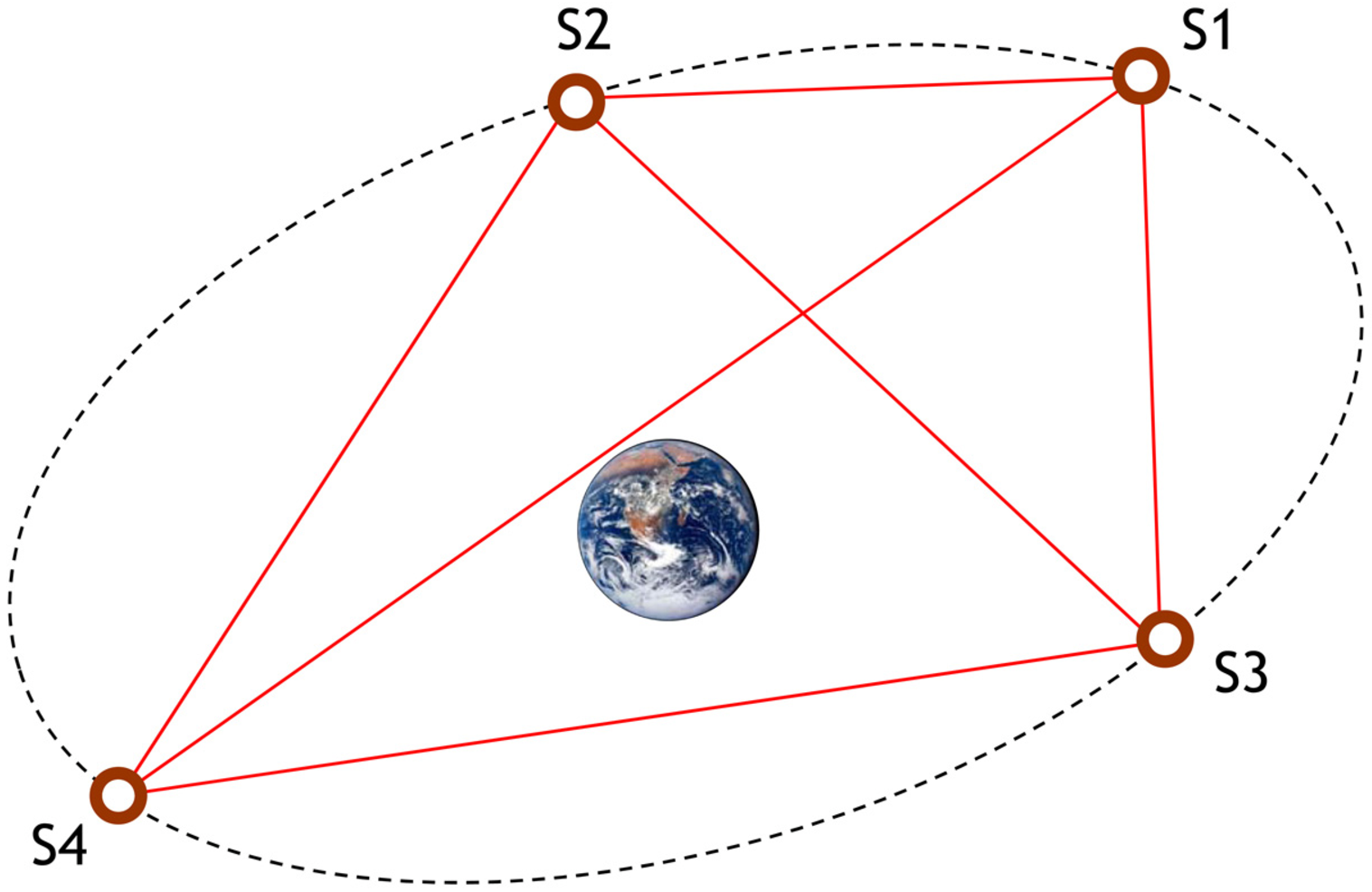}
  \end{center}
  \vspace{-10pt}
  \caption{Schematic of the BEACON formation.}
\label{fig:BEACON-formation}
  \vspace{-15pt}
\end{wrapfigure}

The BEACON mission architecture is based on a formation of four small spacecraft placed in circular Earth orbit at a radius of $\sim$80,000~km (Fig.~\ref{fig:BEACON-formation}). Each spacecraft is equipped with three laser transceivers in order to measure the distances to the other spacecraft in the formation. These laser measurements form a trapezoid with diagonal elements such that one of the legs in the trapezoid skims close to the Earth, picking up an additional gravitational delay; the magnitude of this signal is modulated by moving the position of one of the spacecraft relative to the others (thus changing the impact parameter of the trapezoid legs).  This paper discusses the various aspects of  this mission.

The paper is organized as follows:  Section \ref{sec:beacon} provides the overview for the BEACON experiment, including the preliminary mission design. Section \ref{sec:beacon_instrument} discusses BEACON's current instrument design. In Section \ref{sec:flight-system} we present the mission's flight system design. Section~\ref{sec:error-budget} gives an overview of the preliminary measurement error budget. We present our conclusions in Section~\ref{sec:conclusion}.

\section{The BEACON Concept}
\label{sec:beacon}

The superior sensitivity of BEACON is enabled by a redundant optical-truss architecture that eliminates the need for expensive drag-free systems.  The mission uses four spacecraft that are placed in high-altitude circular orbits around the Earth (Fig.~\ref{fig:BEACON-formation}).  All spacecraft are in the same plane and could potentially be separated by distances up to 160,000 km.  Each spacecraft is equipped with three sets of identical laser ranging transceivers (LRTs) that are used to measure distances between the spacecraft to high accuracy ($\sim$0.1 nm).  In Euclidian geometry this system is redundant; by measuring only five of the six distances one can compute the sixth one.  However, to enable its primary science objective, BEACON will precisely measure and monitor all six inter-spacecraft distances within the trapezoid.

The resulting geometric redundancy is the key element that enables BEACON to measure a departure from Euclidian geometry.  In the vicinity of the Earth, such a departure is primarily due to the curvature of the relativistic space-time around the Earth. It amounts to $\sim$10 cm for light rays that just graze the surface of the Earth and then falls off inversely proportional to the impact parameter.  During a measurement cycle the impact parameter is slowly changed; simultaneous analysis of the resulting time-series of the intersatellite distance measurements will allow BEACON to measure the curvature of the space-time around the Earth.

\subsection{The BEACON Measurement} 
\label{sec:architect}

The experiment will measure changes in the propagation path-length between a pair of spacecraft as the impact parameter (with respect to the Earth) of the beam passing between them is changed. The measurement is performed using a laser metrology system, essentially by counting interference fringes between the transmitted beam and a local reference beam.   Such metrology systems have been developed for the Space Interferometry Mission (SIM) and have achieved accuracies in the single pico-meter range in laboratory tests. For BEACON the desired accuracy goal is comparatively relaxed: 0.1~nm in $\sim$10$^4$ seconds over a distance of 160,000~km. A basic link budget indicates ample signal-to-noise available to the laser link (see Table~\ref{tab:link}) the primary challenge is in systematic instrumental stability rather than statistical measurement noise (see discussion in Sec.~\ref{sec:flight-system}).

\begin{wraptable}{R}{0.37\textwidth}
\vskip -10pt
\caption{BEACON link budget. \label{tab:link}}
\vskip 5pt
\begin{tabular}{|r|l|} \hline  
{\bf Parameter} &  {\bf Value} \\  \hline \hline 
  &\\[-8pt]
Telescope Diameter & 0.1 m \\
Transmitted Power  & 0.1 W \\
Wavelength  & 1064 nm \\
Transmitted Flux&  $5.38\times10^{16}$~ph/s\\
Distance  & 160,000 km\\
Progagation Factor & $2.39\times 10^{-9}$  \\
Transmission Eff. &  9.2\%  \\  
Received Power  & $2.19\times 10^{-12}$~W \\
Photon flux  & $1.18\times10^{8}$~ph/s\\ 
SNR in 1 sec & 1087\\ 
Path $\sigma$ in 1 sec & 0.1~nm \\\hline  
\end{tabular} 
\vskip -5pt
\end{wraptable}

One needs to know the relative positions of the two spacecraft involved in the measurement to an equivalent accuracy.  This is achieved by making use of an ``optical truss'' formed by flying a total of four spacecraft in a planar formation, and measuring the six distances between them.  As long as the formation remains planar, the relative spacecraft positions are fully specified by five parameters (i.e. five of the six measured inter-spacecraft distances).  In other words, one uses five measured inter-spacecraft distances to predict what the sixth distance should be; the difference between the Euclidean prediction and the measured value for the sixth distance is the relativistic signal. Two important challenges are apparent: 1) the formation must remain planar and 2) the locations of the spacecraft must be known with respect to the center-of-mass of the Earth. Both challenges result in a requirement on spacecraft position knowledge to $\sim$1 cm, and position control to $\sim$10 cm.  

\subsubsection{The Orbit} 

The orbit of the four spacecraft is shown in Fig.~\ref{fig:BEACON-formation}.  The orbit chosen for the BEACON formation must satisfy several constraints: the variation in the scientific observable (the relativistic delay) should be maximized, the range rates between the spacecraft should be kept below approximately 10 m/s in order to avoid saturating the distance measuring system, perturbations on the planarity of the formation should be minimized, the required measurement distances should be minimized, and lastly the total $\Delta V$ required to reach and maintain the formation orbit should be minimized.

In order to satisfy all of these opposing constraints we have selected a circular orbit with a semi-major axis in the 80,000 km range, corresponding to a $\sim$2.6 day orbital period.  The spacecraft are spread out along this orbit so as to form a cross (Fig.~\ref{fig:BEACON-formation}); initial true anomalies for the spacecraft are S1: 0$^\circ$, S2: 315$^\circ$, S3: 45$^\circ$ and S4: 185-195$^\circ$.  In this configuration the S1-S4 line-of-sight (LOS) has the smallest impact parameter with respect to the Earth, and thus is most affected by gravitational effects. 

\subsubsection{The Measurement Cycle} 

The first step in the measurement process is to establish the optical truss, i.e. the laser links between the spacecraft.  Given accurate knowledge of the spacecraft positions from GPS and orbit determination, each spacecraft points its gimbaled laser transmitters at the other spacecraft. The required pointing accuracy is $\sim$5 arcsec, although a raster-scan could be used to rapidly search a larger angle. As soon as the laser ``hits'' the target spacecraft, feedback control can be used to hold the pointing stable.  The feedback comes from the receiving spacecraft accurately measuring the angle and/or intensity of the incoming beam.  It should be noted that there is a trade to be explored here between the transmit/receive aperture size and the pointing requirements.  Alternatively, it is possible to use an optical beacon on each spacecraft as a target for the others to aim at; eliminating the need for inter-spacecraft communication links.

Each scientific measurement cycle begins with one of the spacecraft (S4) performing a pair of small ($\sim$1 m/s) maneuvers in order to lower the semi-major axis of its orbit by $\sim$100 km, while keeping the other orbital parameters fixed. Once S4 is in the lower orbit it will begin to ``catch up'' with the rest of the formation; thus the LOS will shift by a few tenths of a degree per day, with the S1-S4 impact parameter changing accordingly. This drift is small enough that the apparent range rate between the various spacecraft is kept below 10~m/s. Once the LOS has drifted far enough a second pair of burns are performed by S4, raising its orbit to be $\sim$100 km above the formation altitude, thus reversing the direction of the LOS drift. Each cycle takes on the order of 10-30 days and is repeated as long as fuel supplies last.

\subsection{Formation Control} 

\begin{wrapfigure}{R}{0.60\textwidth}
  \vspace{-20pt}
  \begin{center}
    \includegraphics[width=0.60\textwidth]{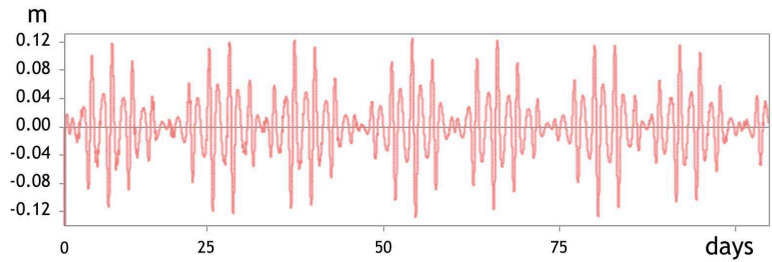}
  \end{center}
  \vspace{-10pt}
  \caption{Out-of-plane error from a simulated formation flight.}
\label{fig:BEACON-z-component}
  \vspace{-10pt}
\end{wrapfigure}

The dominant source of gravitational perturbation in this orbit comes from the Moon; for this reason the orbital plane of the formation is aligned with that of the Moon (limited by lunar precession). If left uncontrolled in the chosen orbit, out-of-plane errors will grow to 100-1000 meters over hour-to-day timescales. However, detailed simulations using a high-precision orbit propagator (including lunar, solar, geopotential and solar radiation pressure models) have shown that even a very simple control system using a milli-Newton-class thruster can reduce the out-of-plane errors to the cm-level (Fig.~\ref{fig:BEACON-z-component}). Small thrusters operate quasi-continuously in the orbit-normal direction to compensate for out-of-plane perturbations. The required $\Delta V$ is a strong function of how closely aligned the formation is to the lunar orbital plane, but for reasonable misalignments, the requirement is $\sim$10-50 m/s/yr.  We have identified a trade between expending propellant to keep the formation plane aligned with the lunar orbit over the duration of the mission (2-3 years), and expending propellant to compensate for out-of-plane errors introduced by the Moon. Based on the simulations run to date we conservatively estimate the total $\Delta V$ required to maintain the formation to be $\sim$200-300 m/s.

Figure~\ref{fig:BEACON-z-component} shows the out-of-plane error for S4. The other satellites are allowed to drift freely in their orbits, while S4 is actively controlled using a PID loop tuned to keep the spacecraft in the plane defined by the other three satellites. The thrust scheme is a continuously adjusted small thrust level, corresponding to accelerations $<3\times 10^{-6}$~m/s$^2$. For a 200~kg spacecraft this can be achieved with a thrust of $\sim$0.6~mN. Total $\Delta V$expended during the 100-day simulation is 5.7~m/s.

\subsubsection{Formation Position Sensing} 

The error knowledge and sensing aspect of the formation flight problem is challenging but tractable. Given that perturbations grow to the tolerance level in a few hours, the control loop must operate above that rate, implying a need for essentially continuous sensing and control. We have identified two possible approaches to solving the problem: 1) continuous measurement of absolute spacecraft position or 2) continuous measurement of spacecraft perturbations coupled with accurate modeling of gravitational effects.

\subsubsection{Position Measurement} 

The first approach requires a means to continuously measure the spacecraft position with an absolute precision in the cm range.  We expect that recent advances in the use of GPS in high orbits (viz. the GSFC Navigator GPS receiver\footnote{The Navigator GPS Receiver, see details at {\tt  http://ipp.gsfc.nasa.gov/ft-tech-GPS-NAVIGATOR.html}}) will enable this approach. Each spacecraft would use a Navigator-type GPS receiver to obtain GPS spillover signals. The acquired pseudo-ranges would be fed to an onboard Kalman-filter to determine the spacecraft position. Brassboard-level hardware-in-the loop simulations indicate that precisions in the sub-meter range are possible. However, if further work shows that the practically achievable signal levels from GPS spillover are insufficient, an alternate approach could be based around ground-based beacon stations (much like existing GPS augmentation systems). These stations would provide pointed high-gain beams aimed at the spacecraft, thus improving the signal strength and resulting positional accuracy. A multi-frequency signal would be required in order to compensate for ionospheric errors, and atmospheric path errors would require precision measurement of dry and wet hydrostatic path terms.

\subsubsection{Perturbation Measurement} 

Given the availability of very high accuracy gravity models (for instance, the GRACE gravity model\footnote{The GRACE gravity model, see {\tt http://www.csr.utexas.edu/grace/gravity/}}), the dominant source of positional error is the uncertainty in the accelerations of the spacecraft due to solar radiation pressure, thruster actuation and other spacecraft events. Hence it may be possible to propagate position models using gravity models and only measure non-gravitational perturbations using highly sensitive accelerometers. On longer time-scales one would rely on ground-based satellite laser-ranging (SLR) for an absolute positional measurement. Assuming laser ranging measurement could be obtained once per day, the required accuracy of the accelerometer would be at the level of just below $\sim 1\times 10^{-9}$~m/s$^2$, the requirement that can be achieved with several types of commercially available accelerometers (for instance,  built by the French company ONERA\footnote{For more details, please see: {\tt http://www.onera.fr/}}).

\subsection{Mission Operations} 

Once the spacecraft have been placed in the science orbit they begin autonomous orbit determination using onboard GPS data.  When formation control is enabled the spacecraft apply low levels of thrust to stay in the desired target plane. 
Data rates will be modest as each laser link may generate data at a rate of less than 0.1 kbps. The largest source of data will be instrumental and environmental health and status information as well as GPS-related navigational data.  Ground control operations will primarily be related to spacecraft health and status monitoring. In addition there will be some active control during the start and stopping phases of each measurement cycle.  A dedicated ground-based laser-ranging effort will be required to determine the absolute craft positions during each measurement cycle. 

\section{BEACON Instrument Description} 
\label{sec:beacon_instrument}

The BEACON instrument consists of three gimbaled laser metrology transceivers, and a camera system to control the pointing. We discuss these systems below:

The laser metrology transceiver consists of the laser, frequency modulators, optics, and frequency stabilizer. The laser light is frequency-stabilized to better than $\Delta\nu/\nu\sim10^{-10}$.  The laser light is then frequency-modulated in order to produce the heterodyne signal and distinguish between incoming and outgoing beams.  Finally, light is collimated and injected into the beam launcher optics. The incoming metrology signal is received by the beam launcher optics and is interfered with the local laser.  A retroreflector serves as the spacecraft fiducial and is common to all three beam launchers.

A 100 mW Nd:YAG laser operating at 1064 nm is used to transmit the metrology signals to the other spacecraft. The laser will be thermally tunable over a range of several GHz. Two lasers are used in each spacecraft for redundancy. The source laser is stabilized to a part in $10^{13}$ short term (this stability has been demonstrated, see Refs.~\citenum{Ye-etal:2003,Foreman-etal:2007a,Foreman-etal:2007b}) using a temperature controlled Fabry Perot etalon. A Pound-Drever scheme is used to servo the frequency of the laser to one of the longitudinal modes of the cavity. Control of the $\sim$3 cm cavity to 10~$\mu$K/s will achieve the required stability.

Acousto-optic modulators (AOM) with fiber-coupled input and output are used to modulate the laser. For a single metrology channel 3 different frequencies are needed for the reference, and two unknown signals. One implementation is a fiber-fed modulator which uses a bulk AOM and is insensitive to alignment errors. Other implementations for the AOMs are possible: these include integrated optic AOMs and multi-channel Bragg cells, both of which will be capable of generating the multiple signals at much lower mass.

The metrology system phase-locks the outgoing laser with the incoming laser. The AOM provides the phase modulation to the laser beam. The incoming signal and the laser output from the AOM are interfered on a high frequency detector. This signal is then used to servo the frequency of the AOM to null. This produces a phase-locked signal whose phase error is determined by the level of the null. In reality, because of the AOM center frequency, the interfered signal will be up-shifted by a stable LO and the servoing done in RF. The stability of this local oscillator is the same as the required stability of the phase locked loop, 10$^{-10}$.  

\begin{wraptable}{R}{0.37\textwidth}
\vskip -10pt
\caption{BEACON spacecraft masses including 30\% built-in contingency. 
\label{tab:masses}}
\vskip 5pt
\begin{tabular}{|l|c|} \hline  
Spacecraft Subsystem &   Design\\ 
 & mass, kg\\  \hline \hline 
Mechanical/Structure    & 39\\
Thermal 	      & 6.5\\
Attitude control  & 9.1\\
RF communications  & 6.5\\
Command/data handling  & 5.2\\
Electric power  & 19.5\\
Propulsion  & 13\\
Harness  & 10.4\\
~~~~Spacecraft bus mass:  & \underline{109.2}\\
Laser local oscillator  & 3.9\\
Beam launchers  & 39\\
GPS receiver/transponder  & 6.5\\
Instrum. thermal control  & 13\\
Instrument harness  & 6.5\\
~~~~Instrument mass:  & \underline{68.9}\\
~~~~Total dry mass:  & \underline{\underline{178.1}}\\
\multicolumn{2}{|c|}{Desired Delta-V: 400 m/s (lsp 224 sec)}\\
Required propellant mass  & \\
Propellant mass provided  & 39.35\\
~~~~Observatory wet mass: & \underline{\underline{217.4}}\\
\hline  
\end{tabular} 
\vskip -35pt
\end{wraptable}

In order to measure the path length to better than $\sim$0.1 nm, errors due to thermal effects on the beam launcher optics must be controlled to the level of 10 mK; consequently an active thermal controller would be used. Furthermore, baffles on the optics will be used to prevent external radiation from affecting the temperature of the instrument.

The acquisition camera will be used as the sensor for pointing the metrology beam.  Three cameras will be used to track each of the incoming metrology beams. In the current instrument design the entire beam launcher optical assembly is gimbaled to point the metrology beam to the target spacecraft.  The 1-axis, 10$^\circ$ range gimbal is used for coarse pointing (all the s/c are co-planar); while a 2-axis, 0.2 arcsec resolution fast-steering mirror is used for fine pointing.

A ground-based time-of-flight laser ranging system is used to triangulate the spacecraft positions with an accuracy of $\sim$1 cm by integrating over a number of laser pulses. A passive corner-cube array is mounted on each spacecraft for this purpose. 

\section{BEACON Flight System} 
\label{sec:flight-system}

The BEACON flight system consists of four spacecraft, which transmit and receive the laser signals to enable the science measurements.  The flight system is subdivided into the instrument payload and the spacecraft bus.
A preliminary mass budget has been established (Table~\ref{tab:masses}).  Conservative estimates for the instrument package on each spacecraft indicate a mass in the 50-55 kg range. Based on commercially available spacecraft busses we estimate the dry mass of a suitable bus to be in the 80-100 kg range. We adopt a contingency factor of 30\%, and include sufficient propellant for 400 m/s of maneuvers to arrive at a nominal spacecraft wet mass of 177 kg.

The carrier structure mass is estimated at 20\% of the carried mass, consistent with experience from the THEMIS mission. We calculate the mass of the propulsion system and propellant required for the $\sim$1400 m/s apogee circularization burn. Including 30\% contingency factors the total flight system wet mass is $\sim$1700 kg.  The Atlas V 501 launch vehicle is capable of injecting $\sim$3600 kg into the desired 80,000 x 167 km orbit, yielding a very comfortable mass margin.

\section{Preliminary Measurement Error Budget} 
\label{sec:error-budget}

The technologies required by BEACON are all rather well-developed, as we explain below. The basic measurement is a path-length between two spacecraft that will be $\sim$160,000 km, with an accuracy goal of 0.1 nm after $\sim10^4$ sec of integration. We have estimated the magnitudes of various terms:

\begin{itemize}
  \item[ i).] Photon Noise:  A simple photon noise budget is provided in Table~\ref{tab:link}.  We assume a diffraction-limited 10-cm diameter transmit/receive aperture, a 100 mW radiated power laser, and an overall 9\% throughput (including the effect of three passes through 50/50 beamsplitters.) The resulting detected photon rate is $\sim 1\times 10^8$ photons/sec; yielding an SNR $\sim10^4$ in one sec, and a corresponding path-length error of order $\lambda/10^4$ = 0.1 nm in one second (see Table~\ref{tab:link}).  

  \item[ii).] Geometric uncertainty of the impact parameter: The desired measurement accuracy is $10^{-9}$; hence the impact parameter should be known to $\sim10^{-9}R_{\rm Earth}$ or 6 mm.  This sets the required knowledge positional accuracy of the spacecraft.

  \item[iii).] Out-of-plane errors:  As a spacecraft drifts ``above'' or ``below'' the plane formed by the other three spacecraft an additional path-length is introduced; the magnitude of this effect is proportional to the cosine of the out-of-plane angle. A 0.1 nm tolerance over 160,000 km corresponds to an angle tolerance $7 \times 10^{-10}$ rad. The equivalent out-of-plane error is 12 cm. This error should be controlled in real time.

  \item[iv).] Optical path changes in the instrument:  The distance measurements between the spacecraft will be made with respect to high-stability optical retro-reflectors onboard each spacecraft. There remain portions of the beam-path where spurious path can be introduced.  The tolerance for such errors of 0.1 nm is achievable with commercially available low CTE optics and overall thermal control at the 10-50 mK level.

  \item[ v).]  Laser Oscillator Stability: The BEACON experiment needs to measure changes to an 80,000km path with $\sim$0.1~nm precision. Nominally this is a measurement to $1 \times 10^{-18}$.  The goal is to measure the Shapiro delay to a precision of $1 \times 10^{-9}$. If we were to use the same laser to measure all 6 optical paths, the requirement for laser frequency stability would be $1 \times  10^{-9}$.  This is easily met with a laser locked to a high finesse Fabry-Perot cavity. The electronic clock used to measure the heterodyne phase also needs to be stable to $1 \times 10^{-9}$, long term. In practice we plan to have a $\sim$100~mW laser on each spacecraft. The light transit time between spacecraft is $\sim$1 sec. Hence on time scales longer than 1 sec we will synchronize the four lasers on the four spacecraft using well established techniques (for details see Refs.~\citenum{Ye-etal:2003,Foreman-etal:2007a,Foreman-etal:2007b}). Note that we only need to measure the frequency differences of the 4 lasers, the frequencies of the four lasers on the four spacecraft do not need to be actively controlled with ultra high precision.

  \item[ vi).]  Frequency Measurement Stability:  The metrology measurement is done by mixing the received laser signal with a local oscillator laser to produce a $\sim$10 MHz tone. The frequency is measured with respect to a temperature-stabilized crystal oscillator with flight heritage.
We have also identified a range of additional sources of error, although order-of-magnitude estimates indicate they are likely smaller in magnitude. The list of issues we have considered includes but is not limited to: beamwalk, vibrations, beam interruptions due to space debris, and the effect of residual atmosphere near Earth.
\end{itemize}

The laser metrology systems at the heart of the BEACON mission arose from developments in the areas of optical astrometry and metrology, primarily as developmental efforts for the SIM and the Keck and Palomar Testbed interferometers.  Thus there now exists brassboard-level development for the laser source, the modulators, the detectors, and the beam optics (including the metrology fiducial multi-axis corner cube.)

The real-time spacecraft navigation will rely on a combination of GPS-based navigation and ground-based laser ranging. High-orbit GPS has been developed for several years with perhaps the most mature hardware being the Navigator receiver.  On-board orbit determination software is at a high level of readiness (viz. the GEONS software package.\footnote{GPS-Enhanced Onboard Navigation System (GEONS), see details at {\tt http://ipp.gsfc.nasa.gov/ft-tech-GEONS.html}})  Ground-based laser ranging now achieves cm-to-mm class distance measurements to the Moon (viz. the APOLLO instrument \cite{Murphy-etal:2007}) on a routine basis.

The spacecraft bus can be a relatively standard commercial-grade platform, with the exception of two areas: low-thrust positional control and low pointing jitter. Attitude knowledge at the 5-10 arcsec level, sufficient for initial laser link acquisition, can be provided by commercially available star trackers.  The fast-steering mirrors needed for jitter stabilization have considerable flight heritage. The position control will require thrusters in the m-Newton range with total velocity increments during the mission being below 50 m/s. Hence cold-gas or mono-propellant systems are usable.

\section{Conclusions} 
\label{sec:conclusion}

Today physics stands at the threshold of major discoveries.  Growing observational evidence points to the need for new physics. 
The recent remarkable progress in observational cosmology has subjected general theory of relativity to increased scrutiny by suggesting a non-Einsteinian model of the universe's evolution. From a theoretical standpoint, the challenge is even stronger -- if gravity is to be quantized, general relativity will have to be modified. Furthermore, recent advances in the scalar-tensor extensions of gravity, brane-world gravitational models, and efforts to modify gravity on large scales motivate new searches for experimental signatures of very small deviations from general relativity on various scales, including on the spacecraft-accessible distances.\cite{Turyshev-etal:2007,Turyshev:2008jd} 

BEACON will lead to very robust advances in the tests of fundamental physics: it could discover a violation or extension of general relativity, and/or reveal the presence of an additional long range interaction in the physical law. With this mission testing theory to several orders of magnitude higher precision, finding a violation of general relativity or discovering a new long range interaction could be one of this era's primary steps forward in fundamental physics. 

Concluding, we would like to emphasize the fact that BEACON is an affordable space-based experiment. 1~W lasers with sufficient frequency stability and over 10 years lifetime have already been developed for optical telecom and also are flight qualified for SIM. Additionally, small optical apertures $\sim$10--20~cm are sufficient and provide this experiment with a high signal-to-noise ratio. This all makes BEACON a scientifically strong and technologically sound candidate for the next flight experiment in fundamental physics for the next decade.  

\section*{ACKNOWLEDGMENTS} 
The work described here was carried out at the Jet Propulsion Laboratory, California Institute of Technology, under a contract with the National Aeronautics and Space Administration, and at the Charles Stark Draper Laboratory, Inc..



\end{document}